\documentstyle[11pt]{article}
\makeatletter
\@addtoreset{equation}{section}
\makeatother

\def\cA{{\cal A}}

\def\cF{{\cal F}}

\def\cH{{\cal H}}

\def\cM{{\cal M}}

\def\cO{{\cal O}}
\def\cP{{\cal P}}

\def\cS{{\cal S}}

\newcommand{\half}{\mbox{\small $\frac{1}{2}$}}

\def\lsim{\;\raisebox{-.4ex}{\rlap{$\sim$}} \raisebox{.4ex}{$<$}\;}
\newcommand{\quarter}{\mbox{\small $\frac{1}{4}$}}
\def\real{\mathop{\rm Re}}

\def\slsh#1{\rlap{$\,/$}#1}
\newcommand{\sixth}{\mbox{\small $\frac{1}{6}$}}

\newcommand{\twentyfourth}{\mbox{\small $\frac{1}{24}$}}

\def\Tr{\mathop{\rm Tr}}

\begin{document}
\begin{titlepage}
\vskip 0.5in
\hfill FERMILAB-PUB-92/133-T
\vskip 1.0in
\begin{center}
{\LARGE Dynamics of Langevin Simulations \par} \vskip 1.5em
{\large \lineskip .5em
\begin{tabular}[t]{c}
Andreas S. Kronfeld \\[2.0em]
Theoretical Physics Group, Fermi National Accelerator Laboratory,\\
P.O. Box 500, Batavia, IL 60510, USA
\end{tabular}\par}
\vskip 1em
{\normalsize \today}
\end{center}
\vfill
\begin{center} \sc Abstract \end{center}
\quotation \normalsize
This chapter reviews numerical simulations of quantum field theories
based on stochastic quantization and the Langevin equation.
The topics discussed include renormalization of finite step-size
algorithms, Fourier acceleration, and the relation of the Langevin
equation to hybrid stochastic algorithms and hybrid Monte Carlo.
\endquotation
\vfill
{\small \noindent
Invited chapter to appear in the special supplement
``Stochastic Quantization'' of {\em Progress of Theoretical Physics}.}
\vfill
\end{titlepage}
\section{Introduction}
A great challenge in particle physics is the numerical
solution of quantum field theories.
Analogous problems appear in condensed matter physics as well.
Stochastic quantization is extremely useful here, because it provides a
direct path from the formal quantization of systems with many degrees of
freedom to practical algorithms for numerical work.
This chapter reviews this connection, leading up to the most popular
algorithms for quantum chromodynamics.

Quantum field theories involve an infinite number of degrees of freedom.
This is the origin of ultraviolet divergencies.
To perform any sensible calculations, the number of degrees of
freedom must be regulated.
This can be done at the level of stochastic differential equations
\cite{Hal92}, but for numerical computations one uses lattice field
theory instead.
The fields of continuous space-time are replaced by aggregate, or
``block'' fields, on the sites or links of a lattice \cite{Wil74}.
For a recent review, see ref.~\cite{Kro92}.
If the lattice has a finite volume, lattice field theory is a quantum
mechanical system with a large but finite number of degrees of freedom.

One way to think of numerical simulations of lattice field theory is as
Monte Carlo integration of the functional integral.
An equivalent way is to imagine integrating the Langevin equations of
stochastic quantization.
The different viewpoints have led to different algorithms; with the
latter approach leading to algorithms based on stochastic difference
equations.
As with deterministic difference equations the general idea is to
devise finite-difference approximations to the differential equations.
However, the random noise affects the analysis of step-size errors
in several ways.

A central focus of this chapter is the dynamics of Langevin simulations.
The term ``dynamics'' does not mean the dynamics of, say, QCD, but the
behavior of the numerical algorithms in simulation, or CPU, time.
In this language, the numerical algorithm is a dynamical system, whose
static behavior is the field theory (e.g.\ QCD) under study.
To solve quantum field theories numerically, it is also important to
analyze the dynamics of the algorithms, because fast dynamics obtain the
solution in less computer time.

This chapter is organized as follows.
Sects.~\ref{Langevin} and \ref{Fokker-Planck} introduce discrete
Langevin and Fokker-Planck equations for scalar field theory.
The analysis is the analog for stochastic differential
equations of the analysis of step-size errors for deterministic
differential equations.
For local field theories renormalization plays an interesting role
\cite{Zin92} and the modifications needed for discrete Langevin
equations are presented in sect.~\ref{universality}.
It is argued that the step-size errors of sects.~\ref{Langevin} and
\ref{Fokker-Planck} do not propagate to physical quantities, when
renormalization is taken into account.
For particle physics applications non-Abelian gauge theories are the
most important systems; they are treated briefly in
sect.~\ref{non-Abelian}.
The analysis of these sections is extended to higher order integration
schemes, such as Runge-Kutta in sect.~\ref{high}.
The so-called hybrid stochastic algorithm is treated here.
Sect.~\ref{fermions} explains algorithms for fermions.
Sect.~\ref{acceleration} discusses several schemes for accelerating the
Langevin dynamics, i.e.\ for generating statistically independent
lattice fields more quickly.
Unfortunately, renormalization does not eliminate step-size errors in
algorithms for fermions or with acceleration, but sect.~\ref{exact}
presents a technique to make the algorithms exact.

Langevin simulations for complex actions are covered in another
chapter \cite{Sch92}.

\section{Discrete Langevin Equations}\label{Langevin}
Consider scalar lattice field theory on a $d$-dimensional hypercubic
lattice with spacing $a$.
The standard action is
\begin{equation}\label{scalar-action}
S = a^d \sum_x \left(- \half\phi_x \triangle \phi_x + V(\phi_x)\right),
\end{equation}
where
$\triangle\phi_x:=\sum_\mu(\phi_{x+\mu}+\phi_{x-\mu}-2\phi_x)/a^2$,
and a typical potential is
\begin{equation}\label{scalar-potential}
V(\phi) = \half m^2 \phi^2 + g \phi^4 .
\end{equation}
The subscripts $x$, etc, denote space-time coordinates.

Stochastic quantization introduces a ``time'' parameter $t$.
The fields evolve in $t$ by a Langevin equation such as
\begin{equation}\label{scalar-Langevin}
\dot{\phi}_i(t) = - \eta_i(t) - \nabla_iS(t),\;\;
\end{equation}
where the dot denotes a derivative with respect to $t$,
the functional derivative
\begin{equation}\label{field-derivative}
\nabla_iS = \frac{1}{a^d} \frac{\partial S\,}{\partial \phi_i} ,
\end{equation}
and $i$ is a multi-index denoting $x$ and any internal indices.
The noise is Gaussian:
\begin{equation}\label{noise-dispersion}
\langle\eta_x(t)\rangle = 0,\;\;
\langle\eta_x(t)\eta_y(u)\rangle = 2\,\delta(t-u)\,a^{-d}\delta_{xy}.
\end{equation}
The lattice spacing has been retained in eqs.~(\ref{field-derivative})
and (\ref{noise-dispersion}) to determine the dimension of $t$ in
eq.~(\ref{scalar-Langevin}).
A scalar field has dimension $[\phi]=\half(d-2)$.
Hence, $[\nabla_iS]=\half(d+2)=[\eta]$, whence $[t]=-2$.

It can be shown that the $t$ average
\begin{equation}
\overline{\cO(\phi)} :=
\lim_{T\rightarrow\infty} \int_0^T dt\,\cO(\phi(t))
\end{equation}
reproduces quantum mechanical vacuum expectation values, i.e.\
\begin{equation}
\overline{\cO(\phi)} =
\frac{1}{Z}\int[d\phi]\,\cO(\phi)\,e^{-S}.
\end{equation}
In the lingo of numerical simulations, eq.~(\ref{scalar-Langevin})
generates $\phi$-field configurations according to
the distribution $e^{-S}$.

In a numerical simulation $t$ corresponds to computer time, but the
computer can only evolve the fields in discrete steps $\varepsilon$.
A discrete approximation to the Langevin requires a finite-difference
prescription for the derivative and a regularization of the Dirac
$\delta$-function.
The Euler scheme is the simplest:
\begin{equation}
\dot{\phi} \mapsto
\frac{1}{\varepsilon}\left(\phi(t+\varepsilon) - \phi(t)\right)
\end{equation}
in eq.~(\ref{scalar-Langevin}), and
\begin{equation}
\delta(t-u) \mapsto \frac{1}{\varepsilon} \delta_{tu}
\end{equation}
in eq.~(\ref{noise-dispersion}).
Higher order schemes will be discussed in sect.~\ref{high}.
Writing $\lambda=t/\varepsilon$ and
$\eta^{(\lambda)}=\eta(t/\varepsilon)/\sqrt{\varepsilon}$ the
Euler update becomes
\begin{equation}\label{update}
\phi_i^{(\lambda+1)} = \phi_i^{(\lambda)} - f_i^{(\lambda)},
\end{equation}
where
\begin{equation}\label{scalar-Euler-force}
f_i^{(\lambda)} =
\sqrt{\varepsilon}\eta_i^{(\lambda)}+\varepsilon\nabla_iS^{(\lambda)}
\end{equation}
and
\begin{equation}
\left\langle\eta_{x,a}^{(\lambda)}\,\eta_{y,b}^{(\mu)}\right\rangle=
2\,\delta^{\lambda\mu}\,a^{-d}\delta_{xy}\delta_{ab}.
\end{equation}

It is instructive to study the Euler algorithm in free field theory,
where it can be solved exactly.
Neglecting internal indices and Fourier-transforming to momentum space
\begin{equation}\label{p-space-solution}
\phi_p^{(\lambda)}=
\left(1-\varepsilon\omega^2(p)\right)^\lambda\phi_p^{(0)} +
\sqrt{\varepsilon}\sum_{\mu=0}^{\lambda-1}
\left(1-\varepsilon\omega^2(p)\right)^{\lambda-\mu-1}\eta_p^{(\mu)},
\end{equation}
where $\omega^2(p)=\hat{p}^2+m^2$, $\hat{p}=2\sin\half p$.
Now consider the correlations in Langevin time, which express the speed
of convergence and de-correlation of the algorithm.
{}From eq.~(\ref{p-space-solution}) one finds
\begin{equation}\label{auto-correlation}
\left\langle \phi_{p_1}^{(\kappa+\lambda)}\,
\phi_{p_2}^{(\lambda)} \right\rangle=
\left(1-\varepsilon\omega^2(p_1)\right)^\kappa
\left\langle\phi_{p_1}^{(\lambda)}\,\phi_{p_2}^{(\lambda)}\right\rangle
\end{equation}
where $\langle \bullet \rangle$ denotes an average over the noise.
Eq.~(\ref{auto-correlation}) reveals many details of the
performance of the Euler algorithm.
First, taking the limit $\varepsilon\rightarrow0$,
$\kappa\rightarrow\infty$ with $t=\varepsilon\kappa$ fixed, one sees
that the convergence and de-correlation of $\phi_p$ occurs in time
$t_{\rm c}(p)=\omega^{-2}(p)$.
Second, for finite $\varepsilon$ one sees that the algorithm is stable
only if $|1-\varepsilon\omega^2(p)|<1$ for all momenta.
The most restrictive mode is the one with the largest momentum
$\omega^2_{\rm max}=4d/a^2+m^2$.
Once $\varepsilon$ is fixed by this ultraviolet mode, the infrared modes
containing the interesting physics de-correlate in
\begin{equation}\label{Euler-slowing-down}
N_{\rm c}(p)=\frac{t_{\rm c}(p)}{\varepsilon}\propto
\frac{4d+a^2m^2}{a^2(\hat{p}^2+m^2)}
\end{equation}
steps of the algorithm.
An important characteristic of simulation algorithms is the critical
dynamical exponent $z$, defined by $N_{\rm c}\propto a^{-z}$ as
$a\rightarrow0$, because the number of steps of the algorithm needed to
completely de-correlate $\phi$ is determined by the largest $N_{\rm c}$.
Typically $z>0$, which means that more and more computation is needed to
simulate field theories as the continuum limit is approached.
This undesirable behavior is called critical slowing down.
{}From eq.~(\ref{Euler-slowing-down}) one sees that $z=2$ for the Euler
algorithm.
It will become clear in sects.~\ref{universality}, \ref{high}, and
\ref{acceleration} that critical slowing down is closely related to the
physical dimension of simulation time.
Except for over-relaxation, cf.\ sect.~\ref{over-relaxation}, the
algorithms considered here all obey $z=-[t]$ (for free field theory).

\section{Discrete Fokker-Planck Equations}\label{Fokker-Planck}
One must now check that the probability distribution is correct as
$\varepsilon\rightarrow0$.
It is also useful to work out the O($\varepsilon$) corrections to the
distribution and develop a formalism for checking that higher-order
schemes are indeed higher order.
To simplify notation this section's equations are in lattice units,
$a=1$.

Let the probability distribution at $\lambda$ be
$\cP^{(\lambda)}[\phi]$.
The update of eq.~(\ref{update}) changes it to
\begin{equation}
\cP^{(\lambda+1)}[\phi]= \int [d\phi']\,
\left\langle\prod_i \delta(\phi_i-\phi'_i+f_i)\right\rangle
\cP^{(\lambda)}[\phi'].
\end{equation}
For small $\varepsilon$ the $\delta$-functions can be expanded in powers
of the drift force $f_i$.
Integrating over $\phi'$ yields the Kramers-Moyal expansion
\begin{equation}\label{Kramers-Moyal}
\cP^{(\lambda+1)}[\phi]= \cP^{(\lambda)}[\phi]+
\sum_{n=1}^\infty \frac{1}{n!}
\nabla_{i_1}\cdots\nabla_{i_n}
\left(\langle f_{i_1}\cdots f_{i_n}\rangle\cP^{(\lambda)}[\phi]\right).
\end{equation}
Eq.~(\ref{Kramers-Moyal}) gives a (functional) differential equation for
the equilibrium distribution.
Working to second order in $\varepsilon$
\begin{equation}\label{Kramers-Moyal-moments}
\begin{array}{r@{\,=\,}l}
\langle f_i \rangle & \varepsilon S_i, \\[0.7em]
\langle f_i f_j \rangle &
2\varepsilon\delta_{ij} + \varepsilon^2 S_i S_j, \\[0.7em]
\langle f_i f_j f_k \rangle &
2 \varepsilon^2 \left(\delta_{ij} S_k +
\delta_{jk} S_i + \delta_{ki} S_j \right), \\[0.7em]
\langle f_i f_j f_k f_l \rangle &
4 \varepsilon^2 \left(\delta_{ij}\delta_{kl} +
\delta_{ik}\delta_{jl} + \delta_{il}\delta_{jk} \right) ,
\end{array}
\end{equation}
using the abbreviation $S_i=\nabla_iS$.

To first order in $\varepsilon$ one obtains the Fokker-Planck equation
\begin{equation}\label{FPE}
\dot{\cP}=\nabla_i[(S_i + \nabla_i)\cP]
\end{equation}
A change of variables $\cP=e^{-S/2}\Psi$ brings eq.~(\ref{FPE}) into
the form
\begin{equation}\label{FPEsa}
\dot{\Psi}= - \left(S_i - \nabla_i\right)
\left(S_i + \nabla_i\right) \Psi=: - \cH \Psi.
\end{equation}
In the space of $\phi$ configurations, $\cH$ is self-adjoint and
positive semi-definite.
The unique zero-mode of $\cH$ is $e^{-S/2}$.
Hence, generic initial conditions converge to the equilibrium solution
$\Psi\propto e^{-S/2}$, or $\cP\propto e^{-S}$.

Now let us analyze the next order in $\varepsilon$.
We are primarily interested in the equilibrium distribution, the
solution to
\begin{equation}\label{order-epsilon}
0= \nabla_i(S_i + \nabla_i)\cP+
\varepsilon\left\{\half \nabla_i \nabla_j(S_iS_j\cP) +
\nabla_i\nabla^2(S_i\cP) + \half\nabla^2\nabla^2\cP\right\} .
\end{equation}
To simplify eq.~(\ref{order-epsilon}) one repeatedly uses
$\nabla_i\cP=-S_i\cP+{\rm O}(\varepsilon)$ inside the braces,
obtaining
\begin{equation}\label{FPE2}
0= \nabla_i[(\bar{S}_i + \nabla_i)\cP],
\end{equation}
where
\begin{equation}\label{equilibrium-action}
\bar{S} = S + \frac{\varepsilon}{4}
\sum_j\left(2S_{jj} - S_j^2\right) .
\end{equation}
Thus, the equilibrium distribution is $\cP\propto e^{-\bar{S}}$, and
$\bar{S}$ is called the equilibrium action.

In field theory the detailed form of the action is not the whole story,
because of renormalization.
Indeed, the terms in the equilibrium action proportional to
$\varepsilon$ are just those appearing in improved lattice actions.
For example, changing field variables
\begin{equation}\label{field-shift}
\phi_i = \tilde{\phi}_i + \quarter\varepsilon S_i
\end{equation}
changes the measure to
\begin{equation}
[d\phi]\simeq [d\tilde{\phi}]\,e^{+\varepsilon \sum_j S_{jj}/4}
\end{equation}
and the action to
\begin{equation}
\bar{S}[\phi]\simeq
\bar{S}[\tilde{\phi}] + \frac{\varepsilon}{4}\sum_j S_j^2,
\end{equation}
up to O($\varepsilon$).
Combining these two changes and writing
$[d\phi]\cP[\phi]=[d\tilde{\phi}]\tilde{\cP}[\tilde{\phi}]$,
the probability distribution $\tilde{\cP}\propto e^{-\tilde{S}}$, where
\begin{equation}
\tilde{S}= S + \frac{\varepsilon}{4}\sum_j S_{jj}.
\end{equation}
For polynomial actions this form modifies the bare couplings but induces
no new terms.

In Euclidean lattice field theory the interesting and accessible
quantities are spectra and matrix elements.
Since eq.~(\ref{field-shift}) is a change a variables, it does not
change the spectrum.
Moreover, any multi-linear observable is correct up to second order:
\begin{equation}
\int[d\tilde{\phi}]\,\cO(\tilde{\phi})\,e^{-\tilde{S}}=
\int[d\phi]\,\cO(\phi)\,e^{-\tilde{S}} + {\rm O}(\varepsilon^2),
\end{equation}
since
\begin{equation}
\int[d\phi]\,\cO_x \tilde{S}_x\,e^{-\tilde{S}}=
-\int[d\phi]\,\nabla_x\left(\cO_x \,e^{-\tilde{S}}\right)=0.
\end{equation}
Hence, a trivial shift in bare couplings and a change of variables
suffices to remove O($\varepsilon$) terms from most interesting
observables.

\section{The Detailed-Balance Universality Class}\label{universality}
The results of the previous section suggest that the non-zero
step-size corrections do not affect physical predictions.
Zinn-Justin has proven a theorem demonstrating the formal
renormalizability of stochastic quantization \cite{Zin86,Zin92}.
The key ingredients of the proof are power counting, a BRST invariance,
and a supersymmetry.
This section examines how to adapt the arguments to analyze the non-zero
step-size corrections to all orders.
One finds \cite{Kro86} that BRST invariance still holds, which
guarantees that the time discretization introduces only
irrelevant operators.
Then the critical phenomena of Euler and related algorithms is
universal.
But the supersymmetry does not hold, so the Fokker-Planck equation is
not necessarily integrable.
However, the pattern of supersymmetry violation is simple, and it is
likely restored in the continuum limit.
If so, the universality class includes the continuous-time Langevin
equation.

The mathematical steps are a direct transcription of ref.~\cite{Zin86}.
Let us start with a general form of the dynamics,
\begin{equation}\label{general-dynamics}
\nu_{x,t}^a=\cF_{x,t}^a[\phi].
\end{equation}
The noise has a slightly different normalization than before
\begin{equation}
\langle\nu_{x,t}^a\rangle = 0,\;\;
\langle\nu_{x,t}^a\,\nu_{y,u}^b\rangle =
\epsilon^{-1} \delta_{tu}\,a^{-d} \delta_{xy}\,\delta^{ab},
\end{equation}
and the unit of time has been changed so that the step-size here is
related to that of sects.~\ref{Langevin} and
\ref{Fokker-Planck} by $\epsilon=2\varepsilon$.
The lattice spacing has been restored here, because we shall treat
discrete space-time and discrete Langevin time on a similar footing.

For convenience, let us introduce some notation.
Greek letters will be used as a multi-index for space-time, internal
indices, and Langevin time, viz.\ $\alpha=(x,a,t)$.
For discrete time one must distinguish forward, backward and symmetric
time difference operators,
\begin{equation}
\partial^\pm_t\phi=
\pm\frac{1}{\epsilon}[\phi(t\pm\epsilon) -\phi(t)],
\end{equation}
and
\begin{equation}
\partial^{(s)}_t\phi=
\frac{1}{2\epsilon}[\phi(t+\epsilon)-\phi(t-\epsilon)].
\end{equation}
Note that $\partial^{(s)}_t=\half(\partial^+_t+\partial^-_t)$.
Dot products, matrix products, or repeated Greek letters imply summation
over all elements of the multi-index {\em and\/} multiplication of
the sum by $\epsilon a^d$, e.g.\
\begin{equation}
J\cdot\phi=J_\alpha\phi_\alpha=
\epsilon a^d \sum_\alpha J_\alpha\phi_\alpha.
\end{equation}
By analogy with eq.~(\ref{field-derivative}), the symbol
$\nabla_\alpha=\epsilon^{-1}a^{-d}\partial/\partial\phi^a_x(t)$.
Finally, functional measures in this section are, for example,
$[d\nu]=\prod_\alpha d\nu_\alpha$.

The generating functional for (dynamic) correlators of $\phi$ is
\begin{equation}\label{dynamic-generating-functional}
Z[J]=\int[d\phi][d\nu]\,\delta(\nu-\cF[\phi])\,\det\cM\,
e^{J\cdot\phi -\half\nu\cdot\nu},
\end{equation}
where the matrix $\cM$ is given by
\begin{equation}\label{ghost-matrix}
\cM_{\alpha\beta}=\nabla_\beta\cF_\alpha.
\end{equation}
One represents the $\delta$-function and determinant as functional
integrals:
\begin{equation}\label{Lagrange}
\delta(\nu-\cF)=\int[d\zeta]\,e^{\zeta\cdot(\nu-\cF)}
\end{equation}
with the contour of $\zeta_\alpha$ along $(-i\infty,i\infty)$, and
\begin{equation}\label{ghost}
\det\cM=\int[dc][d\bar{c}]\,e^{\bar{c}\cM c},
\end{equation}
where the ghosts ($c$) and anti-ghosts ($\bar{c}$) anti-commute.
Inserting eqs.~(\ref{Lagrange}) and (\ref{ghost}) into
eq.~(\ref{dynamic-generating-functional}) and performing
the (Gaussian) integral over $\nu$ yields
\begin{equation}\label{dynamic-generating-functional-noise-out}
Z[J]=\int[d\phi][dc][d\bar{c}][d\zeta]\,
e^{-\cS[\phi,c,\bar{c},\zeta] +  J\cdot\phi},
\end{equation}
where the dynamical action
\begin{equation}\label{dynamical-action}
\cS[\phi,c,\bar{c},\zeta]=-\half\zeta\cdot\zeta + \zeta\cdot\cF
-\bar{c}\cM c .
\end{equation}
There is a BRST transformation,
\begin{equation}\label{BRST-transformation}
\begin{array}{r@{\,=\,}l}
\bar{\delta}\phi_\alpha    & \bar{\xi}c_\alpha, \\[0.7em]
\bar{\delta} c_\alpha      & 0 ,\\[0.7em]
\bar{\delta}\bar{c}_\alpha & \bar{\xi}\zeta_\alpha, \\[0.7em]
\bar{\delta}\zeta_\alpha   & 0 ,
\end{array}
\end{equation}
where $\bar{\xi}$ is an infinitesimal Grassman parameter.
The terms of the dynamical action $\cS$ transform as follows:
\begin{equation}\label{BRST-invariance}
\begin{array}{r@{\,=\,}l}
\bar{\delta}\left(-\half\zeta\cdot\zeta\right) & 0 ,\\[0.7em]
\bar{\delta}\left(\zeta\cdot\cF\right) &
   \bar{\xi}\zeta\cM c, \\[0.7em]
\bar{\delta}\left(-\bar{c}\cM c\right) &
  -\bar{\xi}\left(\zeta\cM c
  +\bar{c}_\alpha\,\nabla_\gamma\nabla_\beta\cF_\alpha\,c_\beta c_\gamma
  \right).
\end{array}
\end{equation}
The last term vanishes because $\nabla_\gamma\nabla_\beta\cF_\alpha$ is
symmetric under $\beta\leftrightarrow\gamma$ whereas $c_\beta c_\gamma$
is anti-symmetric.
The two terms $\zeta\cM c$ cancel.
Hence, the dynamical action is BRST-invariant.

For the study of numerical algorithms a useful class of dynamics is
given by
\begin{equation}\label{general-Langevin}
\cF_{i,t} = \tilde{Q}^{-1}_{ij}
\left(\partial^+_t\phi_{x,t}+\half Q^2_{jk}S_{k,t}\right)
\end{equation}
and $\cM$ defined through $\cF$ by eq.~(\ref{ghost-matrix}).
A Fokker-Planck analysis as in sect.~\ref{Fokker-Planck} shows that this
Langevin equation converges to the correct probability distribution,
if $Q^2$ is positive definite and independent of $\phi$, and if
$\lim_{\epsilon\rightarrow0}(Q^2-\tilde{Q}^2)=0$.
But we shall now apply the functional formalism to see what conclusions
can be drawn for non-zero $\epsilon$.

In the special case $\cF_{i,t}=\partial^+_t\phi_{i,t}+\half\nabla_{i} S$
(i.e.\ the Euler scheme) there is an additional approximate symmetry.
For discrete time the additional transformation is
\begin{equation}\label{super-transformation}
\begin{array}{r@{\,=\,}l}
\delta\phi_\alpha    & \bar{c}_\alpha\xi, \\[0.7em]
\delta c_\alpha      &
  \zeta_\alpha\xi - 2\partial_t^{(s)}\phi_\alpha\xi ,\\[0.7em]
\delta\bar{c}_\alpha & 0 , \\[0.7em]
\delta\zeta_\alpha   & 2\partial_t^{(s)}\bar{c}_\alpha\xi,
\end{array}
\end{equation}
for infinitesimal Grassman $\xi$.
The terms of the dynamical action transform under
eq.~(\ref{super-transformation}) as follows:
\begin{equation}\label{super-invariance}
\begin{array}{r@{\,=\,}l}
\delta\left(-\half\zeta\cdot\zeta\right) &
  2\bar{c}\partial_t^{(s)}\zeta\,\xi ,\\[0.7em]
\delta\left(\zeta\cdot\cF\right) &
  (\zeta\cM\bar{c} + 2\partial_t^{(s)}\bar{c}\cdot\cF)\xi , \\[0.7em]
\delta\left(-\bar{c}\cM c\right) &
  (-\bar{c}\cM\zeta + 2\bar{c}\cM\partial_t^{(s)}\phi)\xi.
\end{array}
\end{equation}
Substituting the expressions for $\cF$ and $\cM$ into
eq.~(\ref{super-invariance}) and collecting all terms
\begin{equation}\label{super-residue}
\delta\cS=
2\bar{c}_\alpha\left[S_{\alpha\beta}\partial^{(s)}_t\phi_\beta-
\partial^{(s)}_tS_\alpha\right]\xi.
\end{equation}
In a formal limit of continuous Langevin time the dynamical action is
invariant, because
$\partial_t S_\alpha=S_{\alpha\beta}\partial_t\phi_\beta$,
from Leibniz' rule.
For interacting theories with discrete Langevin time it is not,
but the residue is a ``lattice artifact.''
We shall return to this point after a discussion of renormalization.

Ref.~\cite{Zin86} shows how power counting and the BRST invariance
restrict the structure of the counter-terms.
For the Euler update, where the step-size has (momentum) dimension
$[\epsilon]=-2$ the dynamical fields have the dimensions
$[\zeta]=\half(d+2)$, $[\phi]=\half(d-2)$ (as expected), and
$[\bar{c}]+[c]=d$.
Renormalizability means that counter-terms in $\cS$ of dimension
greater than $d+2$ are not needed, and the BRST symmetry relates
$\zeta$--$\phi$ terms to $\bar{c}$--$c$ ones.
The argumentation can be translated into the language of lattice field
theory as follows.
Any dynamics $\cF$ will have a relevant part of dimension
$\half(d+2)$ and any matrix $\epsilon a^d\cM$ in the ghost action will
have a relevant part of dimension 2.
The BRST symmetry implies that the relevant parts of $\cF$ and $\cM$ are
related by eq.~(\ref{ghost-matrix}).
In other words, there is a whole universality class of Langevin
algorithms with the same critical dynamics.
This universality class includes more sophisticated discretizations of
Langevin time, such as those in sect.~\ref{high}.
In particular, the physics does not depend on $\epsilon$, up to the
stability requirements discussed in sect.~\ref{Langevin}.

This conclusion is, perhaps, more easily digested by the following
heuristic consideration.
Restoring the lattice spacing, the Langevin step-size is
$\epsilon=\bar{\epsilon}a^2$, by dimensional analysis.
(The dimensionless number typed into the computer is $\bar{\epsilon}$.)
Consider a sequence of simulations with fixed $\bar{\epsilon}$ but the
bare couplings of the static system (the model being simulated) tuned
to approach the continuum limit (of space-time).
Since simulation time is marked of in steps of $\bar{\epsilon}a^2$, it
seems to approach its continuum limit too.
Hence, it is reasonable to guess that the non-zero step-size
algorithms belong to the detailed-balance universality class, because
the continuous Langevin equation obeys detailed balance.
This universality class includes the Metropolis algorithm and other
exact, local algorithms.

To prove the conjecture one must verify the probability distribution at
equilibrium.
{}From the non-perturbative proofs that stochastic quantization
converges to the correct probability distribution \cite{Kir84}, one
realizes that the supersymmetry plays an essential role.
Therefore, let us return to the approximate symmetry in
eqs.~(\ref{super-transformation})--(\ref{super-residue}).
Even for Langevin time discrete, it combines with the BRST
transformation to form something like a super-algebra.
The generators $D$ and $\bar{D}$ (of $\delta$ and $\bar{\delta}$)
satisfy $D^2=0$, $\bar{D}^2=0$ and
\begin{equation}\label{super-commutator}
(D\bar{D}+\bar{D}D)\Phi=2\partial^{(s)}_t\Phi,
\end{equation}
where $\Phi$ is $\phi$, $c$, $\bar{c}$, or $\zeta$.
The right-hand-side is a discrete time-translation operator.
Since $\bar{\delta}\cS$ is a lattice artifact, it should be possible to
adapt the approach of ref.~\cite{Gol89} to the supersymmetry of
eqs.~(\ref{BRST-transformation}), (\ref{super-transformation}) and
(\ref{super-commutator}).
(Ref.~\cite{Gol89} proved that supersymmetry could be restored in the
$d=2$ Wess-Zumino model.
The supersymmetry considered here is even simpler.)
Assuming this strategy succeeds, the renormalized continuum limit of the
dynamical theory is supersymmetric, and is the same as with detailed
balance.


\section{Non-Abelian Spin and Gauge Systems}\label{non-Abelian}
In particle physics the most interesting field theories are non-Abelian
gauge theories and chiral models.
On the lattice the fundamental variables are Lie group elements defined
on links (gauge theories) or sites (chiral models) of the lattice.
The analysis of the previous sections can be adapted to
non-Abelian theories.
The crucial ingredient is to define differentiation in the group
manifold in a way consistent with partial integration over Haar measure.
With such a definition of $\nabla_i$, eq.~(\ref{Kramers-Moyal}) still
holds.

We shall concentrate on unitary groups and use the following
conventions:
The anti-Hermitian generators $T^a$ are normalized by
$\Tr(T^aT^b)=-\half\delta^{ab}$.
The structure constants are given by the commutation relations
$[T^a,T^b]=-f^{abc}T^c$.
Let $\omega^a$ be small parameters and write $\omega=\omega^aT^a$.
The derivative is defined by \cite{Dru83}
\begin{equation}\label{differentiate-function}
f(e^\omega U) = f(U) + \omega^a\nabla^a f(U) + {\rm O}(\omega^2),
\end{equation}
where $f(U)$ is any function of the unitary matrix $U$.
The most useful example is
\begin{equation}\label{differentiate-trace}
\nabla^a\Tr(UV) = \Tr(T^aUV),\;\;
\nabla^a\Tr(VU^\dagger) = -\Tr(VU^\dagger T^a),
\end{equation}
where $V$ is independent of $U$.
The derivatives do not commute (the Lie-group manifold is curved),
$[\nabla^a,\nabla^b]=-f^{abc}\nabla^c$.
The commutation relation is especially easy to verify from
eq.~(\ref{differentiate-trace}).

In field theory one must keep track of a collection of unitary-group
degrees of freedom.
The commutation relation reads
\begin{equation}\label{derivative-commutation}
[\nabla^a_{x,\mu},\nabla^b_{y,\nu}]=
-f^{abc}\nabla^c\delta_{xy}\delta_{\mu\nu}.
\end{equation}
for gauge fields, and a similar expression without the labels $\mu,\nu$
for spin fields.
It is convenient to introduce a multi-index $i=(a,x,\mu)$ for gauge
fields and $i=(a,x)$ for spin fields.
In the following we shall concentrate on gauge fields.

A Langevin update for non-Abelian fields is given by
\begin{equation}
U_{x,\mu}^{(\lambda+1)}=e^{-f_{x,\mu}^aT^a} U_{x,\mu}^{(\lambda)}.
\end{equation}
The Euler drift force is
\begin{equation}
f_i=\sqrt{\varepsilon}\eta_i+\varepsilon\nabla_iS,
\end{equation}
where $\eta_i=\eta_{x,\mu}^a$ are Gaussian random numbers with
dispersion 2.

The equilibrium action can be worked out just as in
sect.~\ref{Fokker-Planck}, taking care that the $\nabla_i$ now longer
commute.
This leads to an new O($\varepsilon$) ``correction'' to the equilibrium
action, which now reads
\begin{equation}\label{gauge-equilibrium-action}
\bar{S}[U]= \left(1+\frac{\varepsilon C_A}{12}\right)S[U]+
\frac{\varepsilon}{4}
\sum_j\left\{2\nabla_j^2S[U] -(\nabla_jS[U])^2\right\},
\end{equation}
where $C_A$ is the Casimir invariant of the adjoint representation
($C_A=N$ for SU($N$)).
The $S_j^2$ term can again be absorbed into a change of variables, and
for a simple plaquette action the remaining O($\varepsilon$) terms are
absorbed into the bare couplings \cite{Bat85}.
For example, the shift in $\beta$ of the Wilson action is
$\beta\mapsto\beta [1+\varepsilon(C_A/12-C_F)]$, taking the shift from
eq.~(\ref{gauge-equilibrium-action}) and the change of variables into
account \cite{Bat85}.
Wilson loops are multi-linear, so that, after the change of variables,
their expectations are correct up to O($\varepsilon$).
Alternatively, one can correct a Wilson loop by a factor of
$1+\varepsilon C_F/4$ per link.
Note that this correction factor drops out of a Creutz ratio.

The analysis of sect.~\ref{universality} can also be extended to
(pure) gauge theories.
Once again, because $[\varepsilon]=-2$, one expects that dynamical
systems with different values of $\varepsilon$ belong to the same
universality class, i.e.\ that renormalization washes out non-zero
step-size effects.

\section{Higher Order and ``Hybrid'' Algorithms}\label{high}
For the systems considered so far, the effects of discrete Langevin time
are analogous to lattice artifacts.
However, it is still sometimes desirable to investigate discretizations
that suppress them.
For example, to find a non-trivial fixed point one must investigate the
phase structure of the lattice theory; this may require more precise
control over parameter-space than what the Euler update would allow.
Also, in the following sections we shall investigate dynamics for which
the renormalization theorem does not apply.
For example, to eliminate critical slowing down, it is necessary to
introduce dynamics with dimensionless time and, hence, different power
counting.
Furthermore, fermionic systems are almost always treated by a
``pseudo-fermion'' fields whose interactions with scalar or gauge fields
is non-local and non-polynomial.

This section considers algorithms that are still approximate, but have
smaller O($\varepsilon$) effects.
Exact algorithms are considered in sect.~\ref{exact}.

First, we shall consider the Runge-Kutta algorithm \cite{Hel79,Dru83}.
The new configuration is obtained from the old one by
\begin{equation}\label{Runge-Kutta-update}
U^{(\lambda+1)}_{x,\mu}=e^{-\tilde{f}_{x,\mu}} U^{(\lambda)}_{x,\mu}
\end{equation}
with
\begin{equation}\label{Runge-Kutta-force}
\tilde{f}_i=\sqrt{\varepsilon}\eta_i+
\half\varepsilon(1+\sixth\varepsilon C_A)
\left(S_i^{(\lambda)}+S_i^{(\lambda+1/2)}\right)
\end{equation}
where $S^{(\lambda+1/2)}$ denotes the action evaluated using the
tentative update
\begin{equation}\label{tentative-update}
U^{(\lambda+1/2)}_{x,\mu}=e^{-f_{x,\mu}} U^{(\lambda)}_{x,\mu}
\end{equation}
where $f$ is an Euler update with the {\em same\/} noise as in
$\tilde{f}$.
Expanding $S^{(\lambda+1/2)}$ in powers of $\sqrt{\varepsilon}$ and
working out the changes to eqs.~(\ref{Kramers-Moyal-moments}) and
(\ref{order-epsilon}), one finds the equilibrium action coincides with
the desired action up to terms of O($\varepsilon^2$).

Another update with O($\varepsilon^2$) accuracy is obtained by
eq.~(\ref{Runge-Kutta-update}) with
\begin{equation}\label{noisy-force}
\tilde{f}_i=\sqrt{\varepsilon}\eta_i+
\varepsilon (1+\twentyfourth\varepsilon C_A) S_i
-\quarter \varepsilon^{3/2} \sum_j S_{ij} \eta_j
\end{equation}
and no tentative update.
For scalar or pure gauge theories eq.~(\ref{noisy-force}) has no
advantage over the Runge-Kutta scheme, which is easier to implement.
However, the generalization when fermions are coupled in saves an
expensive matrix inversion at each step.

The standard hybrid stochastic algorithm can also be considered as an
improved discretization of the Langevin equation.
Consider the following update steps for scalar field theory:
\begin{equation}\label{refreshment}
\left.
\begin{array}{r@{\,=\,}l}
\pi_i^{(\lambda,1/2)}&
\pi_i^{(\lambda,0)} - \half \delta S_i^{(\lambda,0)} \\[0.7em]
\phi_i^{(\lambda,1)}&
\phi_i^{(\lambda,0)} + \delta \pi_i^{(\lambda,1/2)}
\end{array}
\right\},
\end{equation}
where $\pi^{(\lambda,0)}$ is Gaussian noise with unit dispersion,
followed by a molecular dynamics \cite{Cal82} trajectory of $N-1$
steps of
\begin{equation}\label{leap-frog}
\left.
\begin{array}{r@{\,=\,}l}
\pi_i^{(\lambda,n+1/2)}&
\pi_i^{(\lambda,n-1/2)} - \delta S_i^{(\lambda,n)} \\[0.7em]
\phi_i^{(\lambda,n+1)}&
\phi_i^{(\lambda,n)} + \delta \pi_i^{(\lambda,n+1/2)}
\end{array}
\right\}.
\end{equation}
Finally, the new configuration is given by
\begin{equation}\label{hybrid-update}
\phi_i^{(\lambda+1,0)} = \phi_i^{(\lambda,N)} .
\end{equation}
Eq.~(\ref{leap-frog}) is the leap-frog scheme for integrating Hamilton's
equations for Hamiltonian $H=\half\sum_i\pi_i^2+S[\phi]$ through the
$(\pi,\phi)$ phase space; eliminating $\pi$ it is equivalent to the
Verlet scheme for integrating a second-order ordinary differential
equation.
Eq.~(\ref{refreshment}) is the occasional ``refreshment'' of the
velocities needed to average over momentum.

For $N=1$ the hybrid update collapses to the Euler discretization of the
Langevin equation.
The is the basis of the statement that ``Langevin is a special case of
hybrid.''
For $N=2$ the hybrid update is similar in structure to the Runge-Kutta
update, but instead of O($\varepsilon^2$) accuracy the equilibrium
action is given by eq.~(\ref{equilibrium-action}) with
$\varepsilon=\half\delta^2$.
For arbitrary $N$ the structure is similar to higher-order Runge-Kutta
schemes, but the equilibrium action is the same for all $N$.

Again it is instructive to analyze the performance of these algorithms
in free lattice field theory.
In momentum space the leap-frog iteration can be worked out
\begin{equation}
\phi_p^{(\lambda+1,0)}= \phi_p^{(\lambda,N)}=
\cos_N(N\delta\omega(p))\,\phi_p^{(\lambda,0)} +
\frac{\sin_N(N\delta\omega(p))}{\omega(p)}\,\pi_p^{(\lambda,0)},
\end{equation}
where $\sin_N$ and $\cos_N$ are polynomial approximations of sine and
cosine:
\begin{equation}
\sin_N(\theta)=
\sum_{n=0}^{N-1} (-1)^n \frac{\theta^{2n+1}}{(2n+1)!}\,
\prod_{m=0}^{n-1}1-\frac{(m+1)^2}{N^2} ,
\end{equation}
and
\begin{equation}
\cos_N(\theta)=
\sum_{n=0}^N     (-1)^n \frac{\theta^{2n}}{(2n)!}\,
\prod_{m=0}^{n-1}1-\frac{m^2}{N^2} .
\end{equation}
The auto-correlation function is then
\begin{equation}\label{hybrid-auto-correlation}
\left\langle\phi_{p_1}^{(\kappa+\lambda)}
\phi_{p_2}^{(\lambda)}\right\rangle=
\cos_N^\kappa(N\delta\omega(p_1))\,
\left\langle\phi_{p_1}^{(\lambda)}\phi_{p_2}^{(\lambda)}\right\rangle,
\end{equation}
where $\langle\bullet\rangle$ denotes an average over the every noise
$\pi^{(\lambda,0)}$.

Let us consider two idealized limits.
One is the ``Langevin limit''
\begin{equation}
\begin{array}{llll}
N \mbox{ fixed}, & \kappa \rightarrow \infty, & \delta \rightarrow 0, &
t = \half\kappa N^2\delta^2 \mbox{ fixed},
\end{array}
\end{equation}
and the other is the ``molecular dynamics limit''
\begin{equation}
\begin{array}{llll}
\kappa \mbox{ fixed}, & N \rightarrow \infty, & \delta \rightarrow 0, &
\tau = N\delta \mbox{ fixed}.
\end{array}
\end{equation}
In the Langevin limit,
$\cos^\kappa_N(N\delta\omega)\rightarrow e^{-t\omega^2}$ and
the dynamics de-correlates as any Langevin dynamics.
In particular, $z=2$.
In the molecular dynamics limit
$\cos^\kappa_N(N\delta\omega)\rightarrow\cos^\kappa(\omega\tau)$.
If $\omega\tau$ small this limit differs little from the Langevin limit,
and in particular the computation needed to de-correlate the slow modes
is about as much as with the Euler update.
If one tries to make the trajectory length longer, a problem arises
because there is a spread of frequencies in field theory.
For $\tau>2\pi/\omega_{\rm max}$ almost every choice of $\tau$
coincides with a multiple of $2\pi n/\omega(p)$ for some $p$ and $n$
\cite{Mac89}.
This is a piece of bad luck, because such a mode never de-correlates.
As the physical volume of the system increases the density of modes
increases until $\tau$ has nothing but bad luck at all.
Under these circumstances it is difficult to define $z$ sensibly.

It is safe to say that an important attraction of the hybrid algorithm
was the claim that it had $z=1$ because the time parameter of
the trajectories had (momentum) dimension $[\tau]=[\delta]=-1$.
However, numerical studies have shown that the optimal trajectory length
is $\tau_{\rm opt}\sim\pi/2\omega_{\rm max}$, in accord with the above
remarks.
Therefore, the fast, ultraviolet modes de-correlate in one
trajectory, but the slow, infrared modes de-correlate as in usual
Langevin dynamics.
For the {\em standard\/} hybrid algorithm the short, fixed trajectory
length chosen makes it nothing but an elaborate discretization of the
Langevin equation, with step-size $\varepsilon'=\half\tau^2$.
This step-size is larger than in the Euler algorithm (for equal
step-size error), but nevertheless
the stochastic process has dynamical critical exponent $z=2$.
When step-size errors matter, it is not clear which discretization of
the Langevin equation is preferable, hybrid or Runge-Kutta, when all
aspects of the computation are considered.

The trajectory length can be increased to roughly
$2\pi/\omega_{\rm min}$ if its length varied from trajectory to
trajectory \cite{Mac89}.
Remarkably, this solution is an element of the original hybrid algorithm
\cite{Dua85}, in which the trajectory length $N$ was to be chosen with
probability $(1-P\delta)^N$.
Then, although any given mode has bad luck occasionally, most of the
trajectories de-correlate it.
With variable trajectory length and the option to select any
configuration $\phi^{(\lambda,n)}$ for the ensemble, the stochastic
process is no longer in the detailed-balance universality class.
In particular, the proof of convergence must be modified
\cite{And80,Dua86} and the formalism of sect.~\ref{universality} does
not apply.
Nevertheless, the equilibrium action is still given by
eq.~(\ref{equilibrium-action})
(or eq.~(\ref{gauge-equilibrium-action}) for gauge theories) with
$\varepsilon=\half\delta^2$ \cite{Dua86}.
An individual harmonic oscillator has auto-correlation time
$\tau_c=2/P$, provided $P\leq2\omega$ \cite{Dua85}.
For free field theory it is then easy to see that choosing
$P=2\omega_{\rm min}$ de-correlates all modes in (molecular dynamics)
time $\tau_c=\omega_{\rm min}^{-1}$.
As in standard Langevin the maximum step-size is set by stability of the
ultraviolet modes.
Given this step-size, the number of sweeps needed to de-correlate the
infrared modes is
\begin{equation}
N_{\rm c}=\frac{\tau_c}{\delta}\propto\frac{1}{am}.
\end{equation}
The dynamical critical exponent has been reduced to $z=1$, because for
random trajectory lengths, the parameters can be chosen so that CPU time
{\em is\/} molecular dynamics time.

\section{Including Fermions}\label{fermions}
The previous sections considered systems with bosonic degrees of freedom
only.
This section treats algorithms for fermionic degrees of freedom.

First note that the fermions' part of the action can always be
formulated in a quadratic form.
Then
\begin{equation}
\int[d\psi][d\bar{\psi}]e^{\bar{\psi}M[U]\psi} = \det M[U],
\end{equation}
where $U$ denotes gauge fields interacting with the fermions.
For QCD, $M[U]$ is a lattice version of $\slsh{D}+m$, so it is not real.
The standard remedy is to introduce a second flavor of fermion and
introduce a complex field \cite{WeP81}:
\begin{equation}
\det M[U] \det M[U]= \det M^\dagger[U] \det M[U]=
\int[d\varphi]e^{-S_{\rm pf}},
\end{equation}
where $S_{\rm pf}= \varphi^\dagger(M^\dagger[U] M[U])^{-1}\varphi$, and
$\varphi$ is often called the pseudo-fermion field.
Since $M^\dagger=\gamma_5 M\gamma_5$ the pseudo-fermion action can be
written
\begin{equation}\label{pseudo-fermion-action}
S_{\rm pf}=\varphi^\dagger M_5^{-2}\varphi,
\end{equation}
where $M_5=\gamma_5 M$.
Below these details are less important the the form of the action in
eq.~(\ref{pseudo-fermion-action}), so the subscript 5 will be dropped.

The Euler update for the pseudo-fermion is \cite{Uka85,Bat85}
\begin{equation}\label{fermion-Langevin}
\varphi_i^{(\lambda+1)} =
\left(\delta_{ij} - \varepsilon_\varphi M^{-2}_{ij}\right)
\varphi_j^{(\lambda)} + \sqrt{\varepsilon_\varphi}\xi_i
\end{equation}
where $i$ is a multi-index for space-time, color, and spin indices of
$\varphi$, and the Gaussian noise has dispersion
$\langle\xi^\dagger\xi\rangle=2$.
The drift force of a gauge field coupled to the fermion is augmented by
a new term
\begin{equation}
f_i\mapsto f_i
-\varepsilon\varphi^\dagger M^{-2} (\nabla_iM^{-2}) M^{-2} \varphi .
\end{equation}
Notice that the step-sizes $\varepsilon_\varphi$ and $\varepsilon$
appearing in the fermion and gauge updates need not be the same.

These dynamics suffer from a peculiar critical slowing down.
The eigenvalues of $M^2$ for free Wilson fermions are
$\mu(p)=\sin^2(pa)/a^2+(m+\half a\hat{p}^2)^2$.
The fastest modes in eq.~(\ref{fermion-Langevin}) are the low momentum
modes, $\mu^{-1}(0)=m^{-2}$, and their stability restricts the
magnitude of $\varepsilon_\varphi$.
The auto-correlation length of the high momentum modes is long,
$t_{\rm c}=(m+2d/a)^2$, and consequently they de-correlate in
$N_{\rm c}\propto(am)^{-2}$ sweeps.

However, it is easy to eliminate this critical slowing entirely.
Consider eq.~(\ref{general-Langevin}) with $Q=\tilde{Q}=M$, i.e.\
\begin{equation}
\varphi_i^{(\lambda+1)} =
(1 - \varepsilon_\varphi)\varphi_i^{(\lambda)}
+M_{ij}\sqrt{\varepsilon_\varphi}\xi_j,
\end{equation}
and $\langle\xi^\dagger\xi\rangle=2-\varepsilon_\varphi$.
One can show from the Fokker-Planck equation (or BRST techniques!) that
the equilibrium distribution of $\varphi$ is correct to all orders in
$\varepsilon_\varphi$, with this modification in the noise.
One can even set $\varepsilon_\varphi=1$, in which case the
fermion field de-correlates immediately \cite{Bat85}.
Then the gauge-field drift force is augmented by the bilinear noise
term
\begin{equation}\label{bilinear-noise}
f_i\mapsto f_i-\varepsilon\xi^\dagger\cA_i\xi ,
\end{equation}
where $\cA_i= M^{-1} (\nabla_iM^{-2}) M^{-1}$, and
$\langle\xi^\dagger\xi\rangle=1$.

The bilinear noise algorithm of eq.~(\ref{bilinear-noise}) can be
extended to a higher-order scheme using variations of the Runge-Kutta
technique \cite{Bat86,Kro86a}.
However, there is a complication.
Terms of the form $\langle\xi^\dagger\cA_i\xi\xi^\dagger\cA_j\xi\rangle$
make it impossible to integrate the Fokker-Planck equation.
Unfortunately, the higher-order step-size errors in procedures that
remove the non-integrable terms are proportional to the volume
\cite{Fuk87}.
Hence, even though the error is formally higher-order, the step-size
must be chosen to be smaller.
On the other hand, numerical work \cite{Fuk87} indicates that the
non-integrable terms make only a small contribution to observables, and
it is less harmful to leave them in.

In combining the fermion updates into hybrid algorithms, several
approaches are possible.
One can leave $\varphi=M\xi$ fixed during the gauge-trajectory, update
$\varphi$ for fixed $\xi$, or generate a new $\xi$ at each step of the
trajectory.
For an analysis of these possibilities, see ref.~\cite{Got87}.

\section{Accelerating the Dynamics}\label{acceleration}
In free field theory the original hybrid algorithm ameliorates, but
does not eliminate, critical slowing down.
This section uses the Langevin equation to explore two other ways to
attack the problem, over-relaxation and Fourier acceleration.
In the former case the Langevin equation is used as a pedagogical tool;
most implementations rely on other algorithms.
In the latter, however, stochastic difference equations are essential.

The name of the game is to accelerate the dynamics of the slow modes
and thereby reduce the critical dynamical exponent.
It can be determined analytically in free field theory, but reliable
determinations for strongly interacting systems are extremely
difficult.
For four-dimensional interacting systems, such as QCD, it has not yet
proven feasible to quote $z$ with sensible error estimates.

\subsection{Over-relaxation} \label{over-relaxation}
Let us first consider over-relaxation \cite{Adl81}.
The original formulation and most practical implementations do not look
much like the Langevin equation.
It is, however, possible to re-cast it into this form \cite{Neu87}.
Imagine two harmonic oscillators, i.e.\ action
$S=\half\omega_1^2\phi_1^2+\half\omega_2^2\phi_2^2$.
A Langevin equation with the properties of over-relaxation is
\begin{equation}\label{over-relax}
\left(\!\begin{array}{c}
\dot{\phi}_1 \\ \dot{\phi}_2
\end{array}\!\right)=
-\left(\!\begin{array}{rl}
 \omega_1^2\,\cos\theta & \omega_2^2\,\sin\theta \\
-\omega_1^2\,\sin\theta & \omega_2^2\,\cos\theta
\end{array}\!\right)
\left(\!\begin{array}{c} \phi_1 \\ \phi_2 \end{array}\!\right)+
\sqrt{\cos\theta}
\left(\!\begin{array}{c} \eta_1 \\ \eta_2 \end{array}\!\right);
\end{equation}
the Langevin dynamics couples the two modes together.
More generally,
\begin{equation}
\dot{\phi_i}=
-(\cos\theta\,\delta_{ij}+\sin\theta\,\epsilon_{ij})S_j+
\sqrt{\cos\theta}\,\eta_i,
\end{equation}
so the mode-coupling term drops out of the Fokker-Planck equation.
(Here $\epsilon_{ij}$ is the anti-symmetric tensor.)
Hence, eq.~(\ref{over-relax}) generates configurations with the correct
probability distribution, if $\cos\theta>0$.

The eigenvalues of the matrix in eq.~(\ref{over-relax}) dictate
convergence.
They are
\begin{equation}
\nu_\pm= \half(\omega_1^2+\omega_2^2)\,\cos\theta\pm
\half\sqrt{(\omega_1^2-\omega_2^2)^2-
(\omega_1^2+\omega_2^2)^2\,\sin^2\theta}
\end{equation}
Clearly, if
\begin{equation}\label{theta-critical}
\sin^2\theta \geq
\frac{(\omega_1^2-\omega_2^2)^2}{(\omega_1^2+\omega_2^2)^2}
\end{equation}
the eigenvalues form a complex conjugate pair with
$\real\nu_\pm\lsim\omega_1\omega_2$.
The off-diagonal coupling in eq.~(\ref{over-relax}) accelerates the
slow mode at the expense of decelerating the fast mode.
In free lattice field theory $\omega^2(p)=\hat{p}^2+m^2$, and a typical
strategy couples a mode with momentum $p$ to one with momentum
$P=(\pi+p_1, \pi+p_2, ...)$.
Then $\hat{P}^2=4d/a^2-\hat{p}^2$.
The angle $\theta$ is chosen so that eq.~(\ref{theta-critical}) holds
for all $p$.
For example, choosing $\sin\theta=2d/(2d+a^2m^2)$ one finds
$\tau^{-1}(p)=\real\nu_\pm=(m/a)\sqrt{4d+a^2m^2}$ independent of $p$.
In typical implementations a dimensionless step-size is fixed.
Hence, the de-correlation time measured in sweeps is
$N_{\rm c}\propto (am)^{-1}$.
Like the original hybrid scheme, over-relaxation reduces the dynamical
critical exponent to $z=1$, but does not eliminate critical slowing down
completely.

\subsection{Fourier Acceleration}\label{Fourier-acceleration}
In free lattice field theory Langevin updating can be studied exactly in
momentum space, cf.~sect.~\ref{Langevin}.
This analysis shows why the usual dynamics have $z=2$.
It also suggests a remedy.
In eq.~(\ref{p-space-solution}), instead of taking the step-size
independent of $p$, one could just as well take
$\varepsilon(p)=\bar{\varepsilon}/\omega^2(p)$.
This is Fourier acceleration \cite{Par84,Bat85}.
The natural time step is now $\bar{\varepsilon}$, which is
dimensionless, so (for free field theory) it is easy to see that all
modes de-correlate on the same time scale, and that that time scale is
independent of $a$.
This would eliminate critical slowing down at the theoretical level.
The more relevant standard of success is computation.
Fortunately, the cost of the fast Fourier transform (FFT) algorithm
increases only as $V\log V$.

For interacting theories the central question is the tolerable value
of $\bar{\varepsilon}$.
In position space $\varepsilon$ becomes non-local,
$\varepsilon=\bar{\varepsilon}Q^2_{xy}$, where
\begin{equation}
Q^2_{xy}=\int\frac{d^dp}{(2\pi)^d}\;\frac{e^{ip(x-y)}}{\hat{p}^2+m^2}.
\end{equation}
To leading order in $\bar{\varepsilon}$ the equilibrium action becomes
\cite{Bat85}
\begin{equation}\label{FFT-equilibrium-action}
\bar{S} = S + \frac{\bar{\varepsilon}}{4}
\sum_{i,j}Q^2_{ij}\left(2S_{ij} - S_iS_j\right) .
\end{equation}
These interactions can be made local by introducing a new field
$\zeta$ with kinetic term $\half\zeta(\triangle+m^2)\zeta$ and
$\phi$--$\zeta$ interactions $\zeta_iS_i$ and $\zeta_iS_{ij}\zeta_j$
with couplings proportional to $\bar{\varepsilon}$.
With a local field theory, familiar techniques can be used to predict
the form of $\zeta$ interactions on physical quantities.
(It is difficult to determine their size, except that they are
proportional to $\varepsilon$.)
Once the form is known, the step-size errors can be eliminated by
extrapolating---in essence one takes the continuum limit of Langevin
time explicitly.

The field theoretic analysis can proceed from
eq.~(\ref{FFT-equilibrium-action}), or one can use the formalism of
sect.~\ref{universality}.
Starting from the dynamical action in eq.~(\ref{dynamical-action}), it
is convenient to change variables $\zeta\mapsto\zeta Q^{-1}$ and
$\bar{c}\mapsto\bar{c}Q^{-1}$, to make the theory local.
The algebraic form of the BRST transformation does not change, so it can
be used to derive identities relating different quantities, e.g.\
correlators with and without $\zeta$'s and ghosts.
In particular, the $\zeta$ field has space-time interactions; it is the
same field as in the previous paragraph.

If predictions of the $\varepsilon$ dependence are not reliable enough
to extrapolate, generalizations of the Runge-Kutta method are available
\cite{Bat85}, even when fermions are included \cite{Bat86,Kro86a}.
However, while Runge-Kutta processes render the step-size errors
O($\bar{\varepsilon}^2$), the remaining errors are too cumbersome to
analyze conveniently.

A complication arises for non-Abelian gauge theories.
The eigenvalues of the covariant (lattice) Laplacian are approximately
labeled by momentum only in a smooth gauge.
Consequently, it is necessary to fix the gauge before applying Fourier
acceleration, which makes $Q^2$ implicitly field-dependent.
An alternative, using covariant derivatives in the definition of $Q^2$,
makes it explicitly field-dependent.
Either method changes the Fokker-Planck equation at leading order.
The first two moments in eq.~(\ref{Kramers-Moyal-moments}) become
\begin{equation}\label{FFT-Kramers-Moyal-moments}
\begin{array}{r@{\,=\,}l}
\langle f_i \rangle & \bar{\varepsilon} Q^2_{ij} S_j ,\\[0.7em]
\langle f_i f_j \rangle & 2\bar{\varepsilon} Q^2_{ij} .
\end{array}
\end{equation}
The Fokker-Planck equation is then
\begin{equation}
\dot{\cP}=\nabla_i Q^2_{ij}
\left[ \left(S_j +Q^{-2}_{jk}\nabla_l Q^2_{kl} \right)
+\nabla_j\cP \right]
\end{equation}
which equilibrates to the wrong distribution.
This must be repaired by replacing $S_j$ with
$S_j-\nabla_l\log Q^2_{jl}$ in the drift force.
In practice, the repair is implemented using a stochastic
estimator, cf.\ ref.~\cite{Dav90}.

\section{Exact Algorithms}\label{exact}
For systems with fermions, e.g.\ QCD, the non-zero step-size errors can
alter physical results.
Although the effects can be analyzed, the analysis is not especially
straightforward, because the pseudo-fermions interact non-locally.
For Fourier accelerated algorithms, the step-size errors are easier to
analyze, because the algorithm can be re-cast as a local theory.
In both cases, however, an exact algorithm is desirable.
Even for models where renormalization is thought to wash out step-size
effects, most people would prefer an exact algorithm, at least for
psychological reasons.

There is an exact algorithm, well-suited to QCD, called the hybrid Monte
Carlo \cite{Dua87}.
It is based on the hybrid scheme discussed in sect.~\ref{high}, but
configurations are accepted or rejected according to a Metropolis test.
The secret is to apply the test to $H[\pi,\phi]=\half\pi^2+S[\phi]$,
rather than to $S[\phi]$ alone.
Starting with the configuration $\phi^{(\lambda,0)}$, the steps are as
follows:
\begin{enumerate}
\item
Generate $\pi^{(\lambda,0)}$ from a Gaussian distribution
$e^{-\half\pi^2}$.
\item \label{trajectory}
Carry out eq.~(\ref{refreshment}) and $N-1$ steps of
eq.~(\ref{leap-frog}).
\item
Bring $\pi$ up to the same (molecular dynamics) time as $\phi$:
\begin{equation}
\pi_i^{(\lambda,N)}=
\pi_i^{(\lambda,N-1/2)} - \half \delta S_i^{(\lambda,N)}.
\end{equation}
\item\label{Metropolis}
Make the substitution $\phi^{(\lambda+1,0)}=\phi^{(\lambda,N)}$ with
probability $\min\left(1,e^{-\Delta H}\right)$, where
$\Delta H:=H^{(\lambda,N)}-H^{(\lambda,0)}$; otherwise
$\phi^{(\lambda+1,0)}=\phi^{(\lambda,0)}$ unchanged.
\end{enumerate}
When the process is iterated the configurations labeled
$\phi^{(\lambda,0)}$ have the desired probability distribution
$\cP=e^{-S}$.
As in sect.~\ref{high}, an improvement is to randomize the value of $N$
in step~\ref{trajectory} \cite{Mac89}.

The hybrid Monte Carlo algorithm has become the algorithm of choice in
numerical simulations of full QCD.
As such it warrants a review of its own, but that is beyond the scope of
a set of articles on stochastic quantization.
For a review see ref.~\cite{Ken90}.
To give fair comparison to the other algorithms, however, let us work
out the critical dynamical exponent (for free field theory).
The number of trajectories needed to de-correlate the slow modes is
$N_\tau\propto(\tau_{\rm opt}\omega_{\rm min})^{-2}$ and the amount of
computation in a trajectory is proportional to
$N=\tau_{\rm opt}/\delta$.
Therefore, the total amount of computation needed to de-correlate the
slow modes is
\begin{equation}
N_{\rm c}=N_{\rm t}N\propto
\frac{1}{\tau_{\rm opt}^2\omega_{\rm min}^2}
\frac{\tau_{\rm opt}}{\delta}\propto
\frac{\omega_{\rm max}}{\tau_{\rm opt}\omega_{\rm min}^2},
\end{equation}
since $\delta\omega_{\rm max}\lsim1$ for stability of the molecular
dynamics trajectory.
For a fixed trajectories of length
$\tau_{\rm opt}\sim\omega_{\rm max}^{-1}$, one sees that
$N_{\rm c}\propto (am)^{-2}$, i.e.\ $z=2$.
Similarly, for trajectories of variable length with mean
$\tau_{\rm opt}\sim\omega_{\rm min}^{-1}$, critical slowing down is
less severe, and $z=1$.

Hybrid Monte Carlo has an additional source of slowing down in the
infinite volume limit with lattice spacing fixed.
A leap-frog trajectory drifts off the energy shell by an amount of size
\cite{Cre88} $\Delta H=C\sqrt{V}\delta^2+\half C^2V\delta^4+\cdots$.
Consequently, one must reduce the step-size as $\delta\propto V^{-1/4}$,
otherwise the Metropolis test in step~\ref{Metropolis} rejects almost
every trajectory.
This infinite volume slowing down can be alleviated by higher-order
integration schemes.
Since the number of degrees of freedom increases in the continuum limit
too, this characteristic could affect the values given for $z$ in free
field theory.

For QCD hybrid Monte Carlo, with random trajectories and some of the
other ideas proposed in ref.~\cite{Mac90}, seems to be the algorithm of
the near future.
In practice the approach to the lattice-spacing and volume limits is
restricted by computer memory.
As a four-dimensional theory, QCD has enormous memory demands, so with
only moderate critical slowing down and despite some infinite-volume
slowing down, an appropriately tuned hybrid Monte Carlo should be
adequate.

\section*{Acknowledgements}
I would like to thank Prof.\ Namiki, Prof.\ Okano, and the other editors
of this special edition for inviting me to write this overview.
Fermilab is operated by Universities Research Association Inc.\ under
contract with the U.S. Department of Energy.

\end{document}